\documentclass{iopjournal}
\usepackage{float}
\usepackage[numbers,sort&compress]{natbib}
\begin{document}

\articletype{Paper}

\title{Investigating nucleation-driven phase transitions in neopentyl molecular crystals using infrared thermography and polarised light microscopy}

\author{Frederic Rendell-Bhatti$^{1,*,+}$\orcid{0000-0002-3470-786X}, Vinzent G. Hana$^{1,+}$\orcid{0000-0000-0000-0000}, Csongor Joba$^1$, David Boldrin$^1$\orcid{0000-0003-3833-8341}  and Donald A. MacLaren$^1$\orcid{0000-0003-0641-686X}}

\affil{$^1$SUPA, School of Physics and Astronomy, University of Glasgow, G12 8QQ, Glasgow, UK}

\affil{$^*$Author to whom any correspondence should be addressed.}

\affil{$^+$These authors contributed equally to this work.}

\email{fred.rendell@glasgow.ac.uk}

\keywords{Barocaloric, multi-modal microscopy, solid-solid phase transition, plastic crystal, thermal hysteresis}

\begin{abstract}
Sustainable solid-state refrigerants based on barocaloric materials are often limited by thermal hysteresis associated with supercooling effects. Here, we present imaging methods to investigate and compare thermal behaviour and transition kinetics of the barocaloric molecular crystal neopentyl glycol (NPG) with those of a lightly doped derivative, NPG$_{0.99}$PE$_{0.01}$, which incorporates 1 mol\% pentaerythritol (PE). We use temperature-dependent polarised light (PL) microscopy and infrared (IR) thermography to correlate phase transition kinetics and local heat-flow with the bulk thermodynamic response obtained from calorimetry. We show that the doped system exhibits reduced supercooling and thermal hysteresis, attributed to increased microstructural disorder and an increase in the number of nucleation events. These findings provide insight into the design of low-hysteresis barocaloric materials for high-efficiency solid-state cooling applications.
\end{abstract}

\section{Introduction}

Molecular, or plastic, crystals have recently emerged as a promising class of caloric materials for solid-state heating and cooling due to the large latent heat associated with their solid-solid phase transitions \cite{Manosa2013, Moya2014, Kitanovski2015, Moya2020, Boldrin2021, Lloveras2021, Sun2025}. In particular, orientational order–disorder transitions in small-molecule plastic crystals exhibit significant isothermal entropy changes under modest hydrostatic pressure, via so-called `colossal' barocaloric (BC) effects, making them attractive alternatives to vapour-compression systems \cite{Li2019, Lloveras2019, Aznar2021, Dai2023, Piper2025}. Neopentyl glycol (NPG, C$_5$H$_{12}$O$_2$) is a prototypical example that exhibits very large pressure-driven entropy changes near room temperature \cite{Li2019, Lloveras2019}, stimulating widespread interest in the mechanisms that determine BC performance in molecular crystals \cite{Aznar2020, Li2022, Li2022a, DeOliveira2023, Meijer2023, Rendell-Bhatti2024,  Somodi2024, EscorihuelaSayalero2024, Santos2025, Rendell-Bhatti2025a, Chao2025, Rendell-Bhatti2026, Sanuy2025}. However, despite their promising thermodynamic properties, many molecular crystals, including NPG, suffer from sizeable supercooling effects, leading to deleterious thermal hysteresis in their first-order solid–solid phase transitions. The kinetics associated with the supercooled phase transition have been attributed to nucleation-limited transformation pathways and domain-structure effects \cite{Somodi2024}, and microstructural evolution during cycling \cite{Rendell-Bhatti2024}. Because hysteresis directly impedes the reversibility and efficiency of BC devices, understanding how nucleation density, defect populations, and microstructural disorder govern transition kinetics is essential for designing low-hysteresis BC materials.

NPG forms a molecular crystal where the roughly spherical molecules interact through a three-dimensional hydrogen-bonded network. At ambient pressure, NPG undergoes a first-order solid–solid transition near 314 K, between a low-temperature monoclinic phase (space group $P2_1/c$) and a high-temperature orientationally disordered cubic plastic phase (space group $Fm\bar{3}m$) \cite{Chandra1988, Chandra1993}. The transition is accompanied by a large latent heat ($\Delta H \approx$ 110 kJ kg$^{-1}$) and substantial entropy change ($\Delta S \approx$ 320–400 J kg$^{-1}$ K$^{-1}$), which underpin its colossal barocaloric effect \cite{Chandra1991, Tamarit1994}. This is driven by the breaking of the hydrogen-bond network, when molecular orientations become dynamically disordered \cite{Li2020, Li2022, DeOliveira2023, Rendell-Bhatti2025a}. These thermal changes can be tracked as a temperature change ($\Delta T$) using infrared (IR) thermography as the material goes through its exo- and endothermic phase transitions NPG \cite{Rendell-Bhatti2024}, following measurements in other systems \cite{PalomoDelBarrio2016a, Mailhe2021, Pereira2021, Pereira2023}. Furthermore, in the ordered crystal (OC) monoclinic phase of NPG, the anisotropic molecular arrangement happens to give rise to pronounced optical birefringence that vanishes upon the transformation to the disordered cubic plastic crystal (PC) phase. The sharp change in birefringence has previously been used to provide a sensitive optical contrast for tracking phase evolution under polarised light (PL) microscopy \cite{Somodi2024}. Thus, PL microscopy as a function of temperature can be used to track both microstructural changes and the phase fraction of the OC phase ($\Phi_{\mathrm{OC}}$), and IR thermography provides a direct measurement of local thermal changes.


In this work, we exploit the complementary strengths of temperature-dependent IR thermography and PL microscopy, together with bulk differential scanning calorimetry (DSC), to investigate nucleation-driven phase transitions in pure NPG and a lightly doped derivative with 1 mol\% pentaerythritol (PE), NPG$_{0.99}$PE$_{0.01}$. The motivation for this composition is our recent discovery that the addition of small quantities of PE improve the barocaloric performance of NPG blends \cite{Rendell-Bhatti2026}, but here our focus is more on the additional information that microscopy can provide on the transitional behaviour. By mapping nucleation events, transition wave-front dynamics, microstructural disorder, and reconstructing pseudo-calorimetric curves from imaging data, we directly correlate local kinetic behaviour with bulk thermodynamic response. Our multimodal analysis shows that the doped material exhibits enhanced nucleation density and increased microstructural disorder, resulting in reduced supercooling and hysteresis compared with undoped NPG. These results validate imaging-derived pseudo-calorimetry methods to complement DSC and provide insight for designing low-hysteresis BC molecular crystals.

\section{Methods}
\subsection{Sample preparation}
Powdered samples of NPG and PE (99\% purity, Sigma-Aldrich) were used as received. To prepare the NPG$_{0.99}$PE$_{0.01}$ sample, stoichiometric quantities of NPG and PE were mixed and then fully dissolved in ethanol before being left to recrystallize under ambient conditions for a period of 1 week. The sample was then subjected to thermal treatment consisting of heating to 350 K for 30 minutes and cooling to 265 K for 24 hours, which ensured that the samples were in their low temperature phase. Samples were stored in sealed containers at room temperature. 

\subsection{Calorimetry}
Calorimetry was performed on a TA Instruments DSC 250 at a temperature scanning rate of 2.0 K min$^{-1}$ with sample masses around 10 mg. Onset and endset temperatures ($T_{\mathrm{onset}}$,$T_{\mathrm{endset}}$) were defined using 10\% peak signal on either side of the heat-flow peak. The transition width ($\Delta T_{\mathrm{span}}$) was determined as the difference between $T_{\mathrm{onset}}$ and $T_{\mathrm{endset}}$. Transition hysteresis ($\Delta T_{\mathrm{hys}}$) was calculated as $T_{\mathrm{onset,h}} - T_{\mathrm{onset,c}}$, where the subscripts $\mathrm{h}$ and $\mathrm{c}$ correspond to heating and cooling, respectively. Enthalpy ($\Delta H$) and entropy ($\Delta S$) changes were calculated from calorimetry data according to: 

\begin{equation} \label{enthalpy_calc}
\Delta H = \int \dot{Q} \: \mathrm{d}t ,
\end{equation}

\begin{equation} \label{entropy_calc}
\Delta S = \int \frac{\dot{Q}}{T} \: \mathrm{d}t,
\end{equation}

\noindent where $\dot{Q}$ is the heat flow in W kg$^{-1}$.

\subsection{Polarised light microscopy}
Reflection polarised optical microscopy was performed using a GT Vision GXM L3230 microscope with a Linkam PE120 Peltier heating and cooling stage. A small quantity of sample was dissolved into a minimum amount of ethanol and a few drops were deposited onto a glass slide before being left to recrystallise under ambient conditions. All images were obtained using bright-field reflected polarised white light, and the polarised light analyser was set to obtain the maximum image intensity with the sample at 280 K, before thermal cycling. Polarised light was used due to the fact that NPG displays birefringence in the OC phase and not the PC phase, providing a clear intensity difference during the solid-solid phase transition.

The PL microscopy data focuses on the first thermal cycle, due to the low mass of sample used and high volatility of the plastic crystals. Significant material loss through sublimation was observed over repeated cycling, with the sample mass being significantly depleted after 3-4 cycles. We chose to expose the top surface of the sample to reduce the chance of introducing strain from the interaction of the sample with glass slides as it expands and contracts on heating and cooling.

\subsection{Infrared Thermography}
Samples were prepared for IR thermography by pressing 150 mg of sample in powder form into a pellet with 13 mm diameter, to create a disk of sample $<$ 1 mm in thickness. During thermal cycles, samples were placed on a glass slide and heated between 280 K and 330 K, using the same heating and cooling stage as above, in a low humidity nitrogen environment to stop water condensation on the sample. To study the stochastic nature of the nucleation events, 100 mg samples were imaged sequentially through multiple cooling cycles. Images were obtained using an Optris PI640 IR microscope camera as the samples went through their phase transitions on heating and cooling. Images were acquired every second and difference images were constructed by subtracting subsequent frames from preceding ones.

\subsection{Infrared nucleation detection}
Here we briefly describe the methodology for detecting nucleation events in the samples; a full description of this process can be found in Supplementary Note 1. Automated nucleation detection of the IR thermography data was performed using custom code written in Python. The key steps of this process were first to spatially bin and smooth the IR thermography data to reduce noise, then isolate the  nucleation events by utilising an appropriate threshold, similar to approaches elsewhere \cite{CROCKER1996}. A mask was created around all initial nucleation events and automatically expanded outwards around the nucleation region as the phase transition propagated outwards. This process continued until all pixels in the sample were masked, representing completion of the phase transition.

\section{Results \& Discussion}

\subsection{Calorimetry}
\begin{figure}[H]
 \centering
        \includegraphics[width=0.46\textwidth]{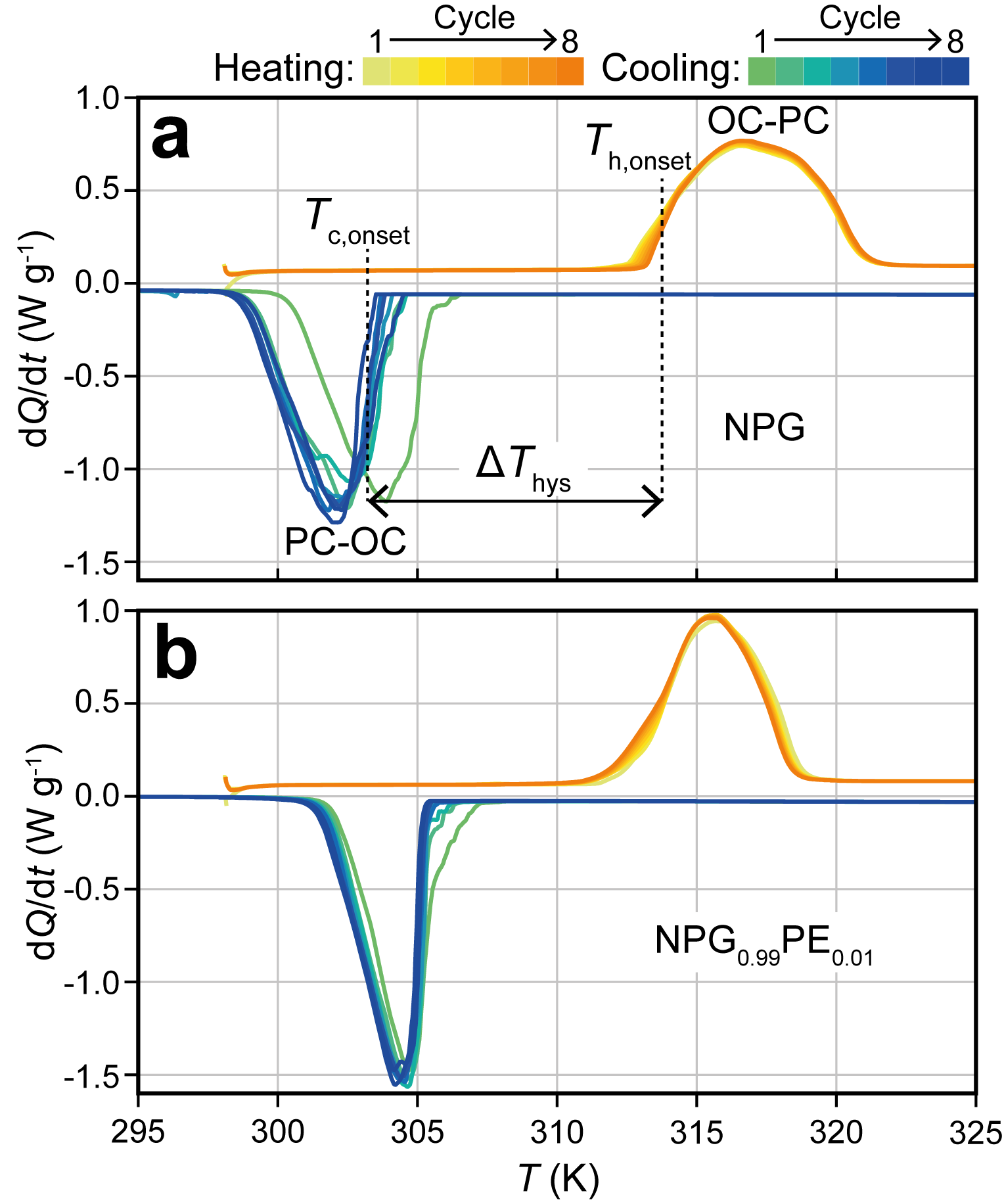}
 \caption{Calorimetry data of NPG and NPG$_{0.99}$PE$_{0.01}$ with a ramp rate of 2 K min$^{-1}$, over eight successive heating and cooling cycles. The method for determining heating and cooling onset and thermal hysteresis is shown in (a). The lightly doped sample of NPG$_{0.99}$PE$_{0.01}$ displays altered phase transition temperatures on both heating and cooling, and greater consistency over successive cycles shown by overlapping heat-flow peaks. }
\label{calorimetry}
\end{figure}

DSC is a common method for characterising phase transitions in caloric materials \cite{Moya2014}. It measures the bulk heat-flow behaviour in a sample during thermal cycles, where sharp peaks (as in Fig. \ref{calorimetry}) signify first-order phase transitions. A characteristic property of molecular crystals is that they exhibit strongly first-order solid-solid phase transitions, often displaying larger latent heat than their corresponding solid-liquid phase transitions. In the neopentyl family of molecules, this is due to the liberation(confinement) of molecular orientational degrees of freedom on heating(cooling), which, in turn, depends on the destruction(formation) of an intermolecular hydrogen bond network. Figure \ref{calorimetry} displays calorimetry data for eight successive heating and cooling cycles of NPG (Fig. \ref{calorimetry}a) and NPG$_{0.99}$PE$_{0.01}$ (Fig. \ref{calorimetry}b), where transitions with $+\frac{dQ}{dt}$ and $-\frac{dQ}{dt}$ correspond to the endothermic OC-PC and exothermic PC-OC solid-solid phase transitions, respectively. The calculated mean transition parameters are given in Table \ref{transition_parameters}, and the full set of parameters can be found in Supplementary Table S1 and S2. 

Comparing Fig. \ref{calorimetry}a and Fig. \ref{calorimetry}b, both materials display hysteresis in their heat flow throughout, as expected. Virgin effects are also observed during the PC-OC phase transition: the initial cycles display slightly different behaviour to that of subsequent cycles. This is more pronounced for NPG, where the first cycle on cooling occurs at a significantly higher temperature than later cycles, perhaps due to microstructural changes on cycling, as observed previously \cite{Rendell-Bhatti2024}. Furthermore, there is significant supercooling exhibited during the PC-OC phase transition in both materials, resulting in a thermal hysteresis (labelled $\Delta T_{\mathrm{hys}}$). However, for NPG$_{0.99}$PE$_{0.01}$ in Fig. \ref{calorimetry}b, the OC-PC phase transition temperature ($T_{\mathrm{h,onset}}$) is reduced by $\sim$1 K and PC-OC phase transition temperature ($T_{\mathrm{c,onset}}$) is increased by $\sim$2 K. The combination of these two effects reduces $T_{\mathrm{hys}}$ by almost 30\%, as shown in Table \ref{transition_parameters}. This is a significant improvement in thermal hysteresis for such a small addition of dopant, and has positive implications for NPG$_{0.99}$PE$_{0.01}$ as a BC refrigerant material \cite{Sun2025}. It is similar to the improvement we recently reported in NPG blends \cite{Rendell-Bhatti2026}. Although $|\Delta S|$ and $|\Delta H|$ are both reduced by around 7\%, the reduced hysteresis will produce a notable improvement to the reversible refrigerant capacity ($\mathrm{RC_{rev}}$) of NPG$_{0.99}$PE$_{0.01}$, as compared to pure NPG.

The DSC data provide rich information on the mean transition parameters, but it is difficult to discern any mechanistic differences between the two materials to account for the reduction of thermal hysteresis. We therefore turn to PL microscopy and IR thermography, since both techniques provide the spatial resolution required to study the phase change to the supercooled state.

\begin{table}[H]
\caption{Transition parameters for NPG and NPG$_{0.99}$PE$_{0.01}$. All parameters are defined in the Methods section and subscripts h and c correspond to heating and cooling, respectively.}
\centering
\begin{tabular}{l c c c c c c}
\hline
Sample & $T_\mathrm{h,onset}$ (K) & $T_\mathrm{c,onset}$ (K) & $\Delta T_\mathrm{hys}$ (K) & $\Delta T_{\mathrm{h,span}}$ (K) & $\Delta H_{\mathrm{h}} $ (kJ/kg)& $\Delta S_{\mathrm{h}} $ (J/kg$\cdot$K) \\
\hline
NPG & 313.0 $\pm$ 0.1 & 304.0 $\pm$ 0.1 & 9.0 $\pm$ 0.1 & 8.0 $\pm$ 0.1 & 113.4 $\pm$ 0.1 & 358.1 $\pm$ 0.4 \\
NPG$_{0.99}$PE$_{0.01}$ & 312.2 $\pm$ 0.1 & 305.6 $\pm$ 0.2 & 6.6 $\pm$ 0.2 & 6.3 $\pm$ 0.1 & 104.8 $\pm$ 0.2 & 332.4 $\pm$ 0.5 \\
\hline
\end{tabular}
\label{transition_parameters}
\end{table}

\subsection{Polarised Light Microscopy}

\begin{figure}[H]
 \centering
  \includegraphics[width=0.7\textwidth]{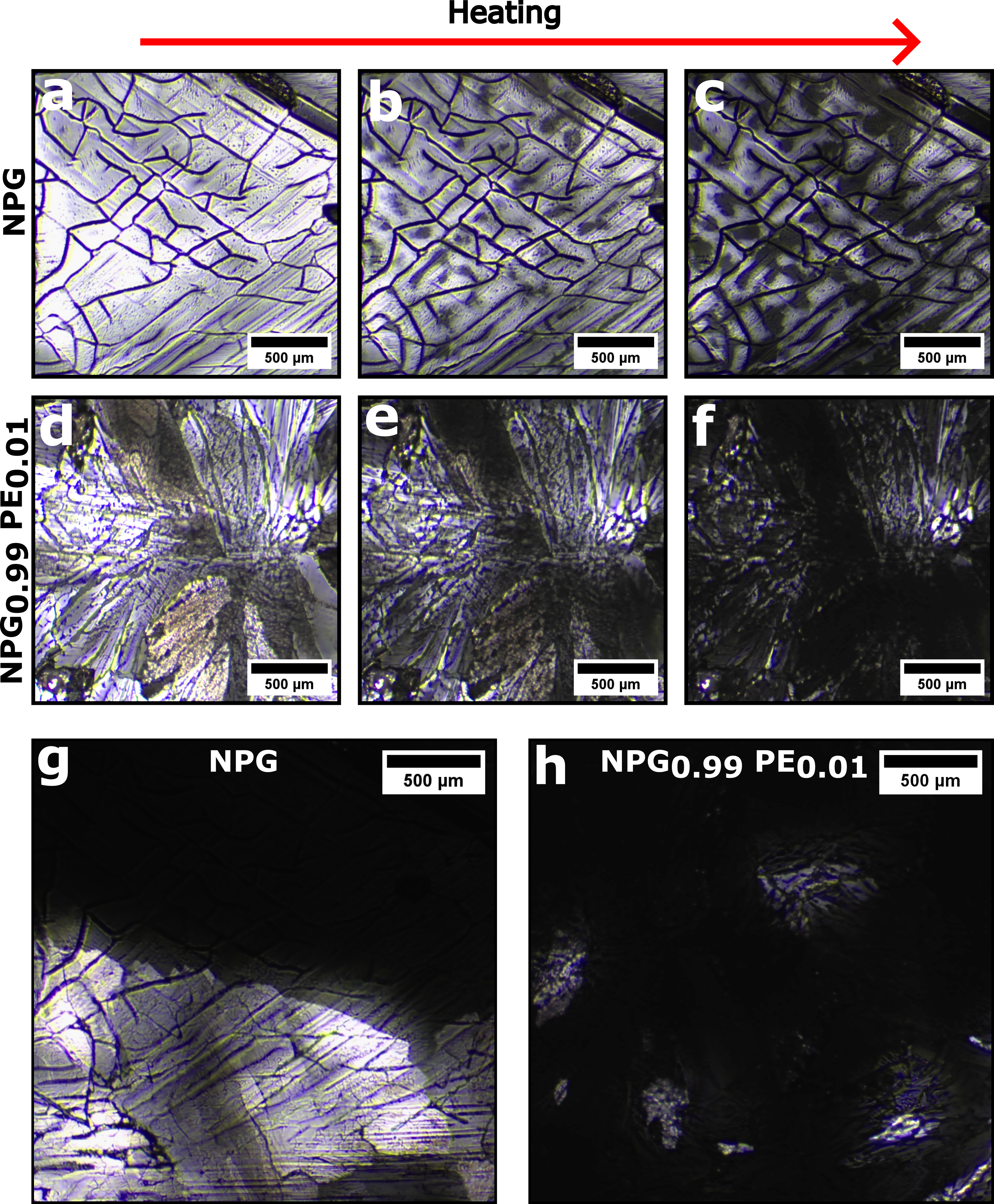}
 \caption{Polarised light microscopy of NPG and NPG$_{0.99}$PE$_{0.01}$ thin layers during their solid-solid phase transition. A series of PL microscopy images capturing the phase transition upon heating for (a-c) NPG and (d-f) NPG$_{0.99}$PE$_{0.01}$. These depict the spatial propagation of the transitions across the field of view. Representative snapshots obtained during the cooling phase transition for (g) NPG and (h) NPG$_{0.99}$PE$_{0.01}$, highlighting the characteristics and quantity of nucleation events between the pure and doped samples.}
\label{PLM_phase_transition}
\end{figure}

Figure \ref{PLM_phase_transition} presents PL microscopy images of thin drop-cast samples of NPG and NPG$_{0.99}$PE$_{0.01}$ during the solid-solid phase transition on heating and cooling. The intensity of the images represents the state of the sample, with high intensity (bright regions) corresponding to the birefringent OC phase and low intensity (dark regions) corresponding to the non-birefringent PC phase. The intensity difference can be spatially tracked as a function of temperature to determine transition behaviour, producing DSC-like curves, as shown in Supplementary Fig. 1. This analysis is consistent with the calorimetry data presented in Fig. \ref{calorimetry} and IR thermography data presented in Fig. \ref{infra-red_comparison}. A more complete set of PL microscopy images during heating and cooling is provided in Supplementary Figs. S2-S5 and the full phase transitions on heating and cooling can be viewed in Supplementary Movies S5-S8.

Figure \ref{PLM_phase_transition}a-c and \ref{PLM_phase_transition}d-f show the OC-PC phase transition upon heating for NPG and NPG$_{0.99}$PE$_{0.01}$, respectively. Initial observations of Fig. \ref{PLM_phase_transition}a and Fig. \ref{PLM_phase_transition}d reveal pronounced structural and topographical differences between the two materials that persist after cycling (Supplementary Figs. 4 and 5). The NPG morphology is characterised by flat platelet regions with dark boundaries corresponding to cracks that have formed during the drop-casting of the sample. NPG$_{0.99}$PE$_{0.01}$ has fewer pronounced cracks and less uniformity of appearance, indicating increased microstructural and crystalline disorder. The crystallites in NPG$_{0.99}$PE$_{0.01}$ appear much smaller and rougher than those in NPG. 

During the heating of NPG (Fig. \ref{PLM_phase_transition}a-c), the phase transition appears to initiate primarily within the platelets, as indicated by the initial loss of PL intensity in these regions (Fig. \ref{PLM_phase_transition}b). This is somewhat surprising, as phase transitions are commonly observed to nucleate at defects, surfaces and interfaces. It could be due to a combination of thermal gradient and mechanical strain effects. If the cracks act as thermal discontinuities, then the central regions will be slightly warmer than the edges and will therefore transition first. In addition, the edges are more free to expand during the OC-PC phase transition, which exhibits a 5 \% volume change, whereas the central regions are not. If the cubic PC phase is able to accommodate strain more isotropically than the monoclinic OC phase then this provides a thermodynamic driving force that favours the PC phase in the central regions. At later stages, the phase transition does seem to also propagate from the cracks (Fig. \ref{PLM_phase_transition}c). While this boundary-led initiation has been previously reported for NPG \cite{Somodi2024}, the observed presence of both intra- and inter-granular transition sites suggests the process is strongly mediated by the sample microstructure.

In contrast, NPG$_{0.99}$PE$_{0.01}$ (Figs. \ref{PLM_phase_transition}d-f) exhibits a more homogeneous phase transition throughout the field of view, where contrast changes appear to be occurring on a smaller length scale, mainly originating from the cracks/grain boundaries. This may indicate lower crystallite sizes and/or more disorder in NPG$_{0.99}$PE$_{0.01}$ as compared to NPG. It is worth noting that the spatial PL contrast variations observed in these samples on heating will not be present during IR thermography (Fig. \ref{infra-red_comparison}b-d,e-g) due to the reduced spatial resolution of the IR images.

Distinct differences in the kinetic behaviour are also observed upon cooling of the two samples in Fig. \ref{PLM_phase_transition}. The region of NPG imaged in Fig. \ref{PLM_phase_transition}g exhibits a single nucleation event and a transformation wave-front that propagates across the field of view (Supplementary Fig. S4), while the doped NPG$_{0.99}$PE$_{0.01}$ sample (Fig. \ref{PLM_phase_transition}h) displays multiple  discrete nucleation events arising in an area of the same size. These differences, during both heating and cooling, suggest a uniform dispersion of the PE dopant throughout the NPG, and a profound influence of a small quantity of dopant on the long range crystalline order and kinetic phase-transition behaviour of the material. 

\subsection{Infra-red Thermography}

\begin{figure}[H]
 \centering
        \includegraphics[width=0.90\textwidth]{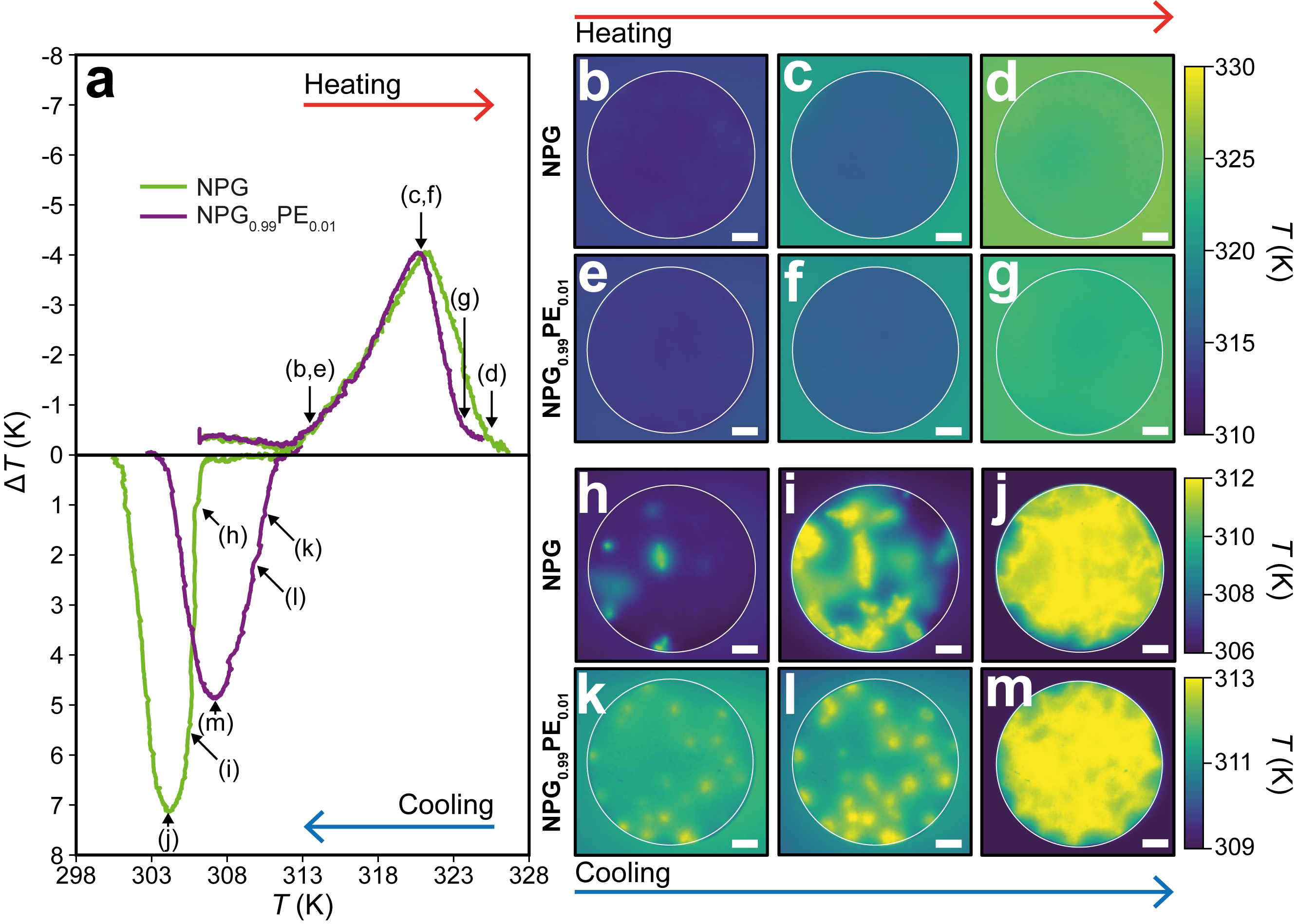}
 \caption{IR images of NPG and NPG$_{0.99}$PE$_{0.01}$ with a ramp rate of $\pm$2 K min$^{-1}$ on heating and cooling. The samples were cycled three times between 280 and 330 K before the images were acquired at temperatures on heating: (b,e) 314 K, (c,f) 321 K, (d) 325 K, (g) 323 K and cooling: (h) 306 K, (i) 305.5 K (j) 304 K, (k) 310.5 K, (l) 310 K, (m) 307.5 K. Both materials exhibit very similar behaviour on heating, where they appear to transition homogenously across the sample. On cooling there are inhomogeneous hot-spots associated with nucleation events and phase transition propagation wave-fronts. The scale bar is 2 mm and circles indicate the sample edge.}
\label{infra-red_comparison}
\end{figure}

Figure \ref{infra-red_comparison} shows IR thermography $\Delta T$ plots (Fig. \ref{infra-red_comparison}a) and images (Fig. \ref{infra-red_comparison}b-m) for NPG and NPG$_{0.99}$PE$_{0.01}$ on heating and cooling during a thermal cycle. By comparing the temperature of the sample pellets (region within the white circle) to the Peltier stage (region outside the white circle), it is possible to track $\Delta T$ of the samples as they go through their solid-solid phase transitions, as shown in Fig. \ref{infra-red_comparison}a. The plot has many similarities to the calorimetry data shown in Fig. \ref{calorimetry}, specifically the reduction in supercooling and thermal hysteresis of NPG$_{0.99}$PE$_{0.01}$ as compared to NPG. It is worth noting that the measured hysteresis is lower in IR thermography than calorimetry for both samples. This is likely because the sample mass is over ten times greater for the IR measurements, which increases the probability of homogenous nucleation during the PC-OC transition \cite{Lilley2021,Shamseddine2022}. Figure \ref{infra-red_comparison}b-d and Fig. \ref{infra-red_comparison}e-g present the IR images obtained at the times labelled in Fig. \ref{infra-red_comparison}a.

Very little difference is observed when comparing the two materials' behaviour on heating in Fig. \ref{infra-red_comparison}b-g, and the OC-PC phase transition appears to be uniform throughout the pellets. In contrast, Fig. \ref{infra-red_comparison}h-j (NPG) and Fig. \ref{infra-red_comparison}k-m (NPG$_{0.99}$PE$_{0.01}$) show a number of discrete hot-spots on cooling through the PC-OC phase transition (full datasets are presented in Supplementary Fig. S6 and S8). These hot-spots correspond to nucleation events and phase transition propagation wave-fronts. This type of nucleation-driven, kinetic behaviour is characteristic of supercooled first-order phase transitions, for example in supercooled water \cite{Li2024}. However, there are some key differences between the samples, particularly related to the number of nucleation events appearing during the onset of the phase transitions. NPG displays fewer nucleation events than those for NPG$_{0.99}$PE$_{0.01}$, and the shapes of the hot-spots corresponding to the transition wave-fronts also appear more anisotropic. Overall, NPG exhibits more elongated heat-flow patterns (Fig. \ref{infra-red_comparison}i) whereas NPG$_{0.99}$PE$_{0.01}$ displays a more speckled pattern (Fig. \ref{infra-red_comparison}l). This variation is even more apparent in the difference IR images presented in Supplementary Figures S7 and S9, where the nucleation events and phase transition wave-fronts are clearly visible as bright regions in the images. The anisotropic heat-flow pattern of NPG mirrors the crystal habit demonstrated in Fig. \ref{PLM_phase_transition}a, and the more isotropic behaviour of NPG$_{0.99}$PE$_{0.01}$ is consistent with the increased microstructural disorder shown in Fig. \ref{PLM_phase_transition}d. Furthermore, the increased width of the cooling transition of NPG$_{0.99}$PE$_{0.01}$ shown in Fig. \ref{infra-red_comparison}a corresponds to the more gradual appearance of nucleation events, and their propagation throughout the sample. This observation is consistent with nucleation rate being proportional to the degree of supercooling \cite{Lilley2021}. The complete phase transitions on heating and cooling of these samples can be viewed in Supplementary Movies S1-S4.

\subsection{Nucleation detection and analysis}

\begin{figure}[H]
 \centering
        \includegraphics[width=0.83\textwidth]{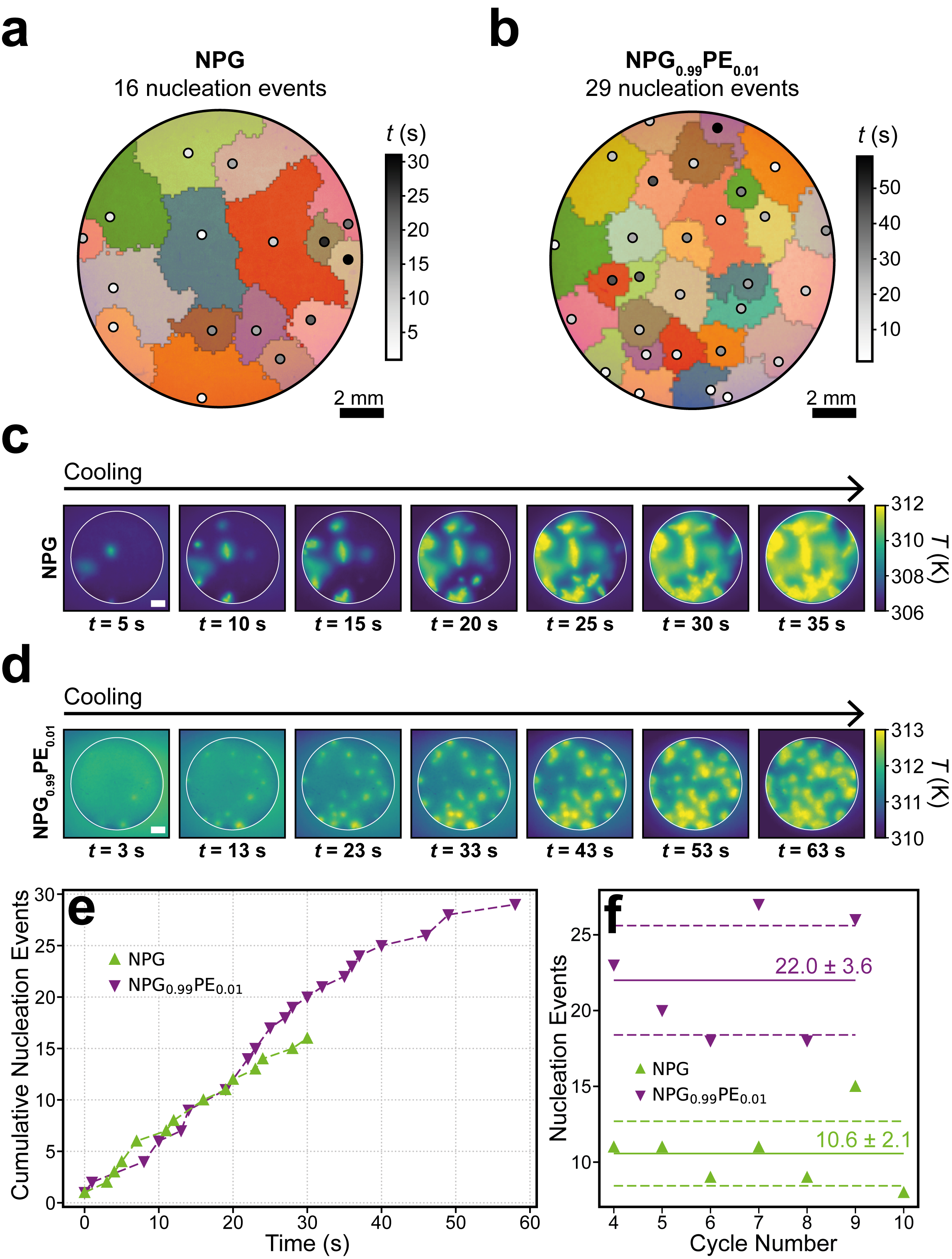}
 \caption{Nucleation event analysis of NPG and NPG$_{0.99}$PE$_{0.01}$ during their supercooled PC-OC phase transitions. Nucleation distribution maps of (a) NPG with 16 total detected nucleation events and (b) NPG$_{0.99}$PE$_{0.01}$ with 29 total detected nucleation events. The greyscale circles denote the initial site of the nucleation, with the coloured surrounding regions corresponding to the spatial extent of each nucleation wave-front propagation. IR image frames through the supercooled phase transition in (c) NPG and (d) NPG$_{0.99}$PE$_{0.01}$, with nucleation events clearly visible as high-temperature regions. (e) Nucleation events as a function of time for the phase transitions of NPG and NPG$_{0.99}$PE$_{0.01}$ shown in (c,d). (f) Nucleation events as a function of cycle number, over ten successive heating and cooling cycles.}
\label{nucleation_pt_analysis}
\end{figure}

Figure \ref{nucleation_pt_analysis} presents the nucleation detection analysis using the IR thermography data presented in Fig. \ref{infra-red_comparison} for NPG and NPG$_{0.99}$PE$_{0.01}$ during the fourth cooling cycle. Figure \ref{nucleation_pt_analysis}a and Fig. \ref{nucleation_pt_analysis}b correspond to the nucleation distribution maps for NPG and NPG$_{0.99}$PE$_{0.01}$, respectively. The introduction of PE as a dopant results in a substantial increase in nucleation site density, as previously reported \cite{Rendell-Bhatti2024}.
In NPG, nucleation events first appear in the left of the image, then later appear at the right. This is indicated by the gradient shading of the circular markers in Figure \ref{nucleation_pt_analysis}a,b. In NPG$_{0.99}$PE$_{0.01}$ (Fig. \ref{nucleation_pt_analysis}b), the nucleation events appear more randomised, consistent with the increased microstructural disorder observed in the PL microscopy data (Fig. \ref{PLM_phase_transition}d). 

Figures \ref{nucleation_pt_analysis}c,d present a series of IR images during the cooling process of the two materials.  It should be noted that not all the thermal hot-spots shown here correspond to nucleation sites as many relate to propagation of the phase transformation away from an earlier nucleation site. We implemented custom code to identify and isolate nucleation sites, taking into account the expected phase transition propagation: details can be found in the Methods and Supplementary Note 1. 

Figure \ref{nucleation_pt_analysis}e shows the timeline of the nucleation events where the samples exhibit a similar initial trend of cumulative nucleation events over time. However, in NPG the total number of nucleation events, and the time over which they occur, is approximately half that of NPG$_{0.99}$PE$_{0.01}$. Thus, the volume of material associated with each nucleation event is greater in NPG, with the phase transformation wave-fronts propagating further. This is also evidenced in the nucleation distribution maps (Fig. \ref{nucleation_pt_analysis}a,b), where the area reached by each nucleation event is much larger in NPG, especially for the first $\sim$5 nucleation events.

Figure \ref{infra-red_comparison}a shows a similar relation with the overall phase transition of the doped NPG$_{0.99}$PE$_{0.01}$ taking longer than pure NPG.
The stability of these nucleation characteristics across multiple heating and cooling cycles is summarised in Figure \ref{nucleation_pt_analysis}f. For these measurements, new pellets were prepared, which were derived from the same powder sample batch. The datasets are presented from the fourth cycle onward to focus on the material performance after it reached a stable microstructural state. The complete dataset is provided in Supplementary Fig. S10. The frequency of nucleation events in both samples remains relatively stable from cycle four onwards. NPG$_{0.99}$PE$_{0.01}$ consistently displays more than double the nucleation events compared to pure NPG. The standard deviation of the nucleation event count is also about 50\% higher in NPG$_{0.99}$PE$_{0.01}$.

\section{Conclusions}
We have presented imaging methods to investigate the thermal behaviour and transition kinetics of the archetypal molecular crystal neopentyl glycol (NPG) and its doped derivative, NPG$_{0.99}$PE$_{0.01}$. We used IR thermography to view the spatial characteristics, and hence the transition kinetics, of the two materials, showing that NPG$_{0.99}$PE$_{0.01}$ exhibits roughly twice the number of nucleation events during its supercooled phase transition. Analysis of the IR data provided pseudo-calorimetric curves that were in good agreement with conventional calorimetry heat-flow data. In addition, the birefringence of NPG also allowed us to exploit PL microscopy to provide structural insight (although other microscopies could provide similar detail). This combined analysis reveals that increased microstructural disorder caused by the dopant also increased the number of nucleation events and explains the improved thermal hysteresis. 

These findings demonstrate a powerful framework for investigating nucleation-limited kinetics of supercooled first-order phase transitions. Here, we have applied this methodology to barocaloric molecular crystals, but it will also be applicable to other solid-state phase change materials.

\ack{This work was financially supported by an EPSRC grant (EP/V042262/1) and The Carnegie Trust Research Incentive Grant (RIG013328).}

\roles{ 

David Boldrin \orcid{0000-0003-3833-8341}

Conceptualization (Equal), Funding acquisition (Equal), Methodology (Equal), Supervision (Equal), Writing - review \& editing (Equal). 

\vspace{0.5cm}

\noindent Donald A. MacLaren\orcid{0000-0003-0641-686X}

Conceptualization (Equal), Funding acquisition (Equal), Methodology (Equal), Supervision (Lead), Writing - review \& editing (Equal). 

\vspace{0.5cm}

\noindent Vinzent Hana \orcid{0000-0000-0000-0000}

Formal analysis (Equal), Investigation (Equal), Methodology (Equal), Visualization (Equal), Writing - original draft (Equal), Writing - review \& editing (Equal). 

\vspace{0.5cm}

\noindent Csongor Joba

Investigation (Supporting), Methodology (Supporting), Writing - review \& editing (Supporting).

\vspace{0.5cm}

\noindent Frederic Rendell-Bhatti \orcid{0000-0002-3470-786X}

Conceptualization (Equal), Formal analysis (Equal), Funding acquisition (Supporting), Investigation (Equal), Methodology (Lead), Supervision (Equal), Visualization (Equal), Writing - original draft (Equal), Writing - review \& editing (Equal).}

\data{Data will be made available on acceptance of manuscript.}

\bibliographystyle{unsrtnat}
\bibliography{Decarbonising_Barocalorics}


\end{document}